\documentclass[twocolumn]{aastex6}
\usepackage{gensymb}
\usepackage{natbib}
\begin{document}
\submitted{}
\title{ALMA Thermal Observations of a Proposed Plume Source Region on Europa}

\author{Samantha K. Trumbo and Michael E. Brown}
\affil{Division of Geological and Planetary Sciences, California Institute of Technology, Pasadena, CA 91125, USA}
\author{Bryan J. Butler}
\affil{National Radio Astronomy Observatory, Socorro, NM 87801, USA}

\begin{abstract}
We present a daytime thermal image of Europa taken with the Atacama Large Millimeter Array. The imaged region includes the area northwest of Pwyll Crater, which is associated with a nighttime thermal excess seen by the Galileo Photopolarimeter Radiometer and with two potential plume detections. We develop a global thermal model of Europa and simulate both the daytime and nighttime thermal emission to determine if the nighttime thermal anomaly is caused by excess endogenic heat flow, as might be expected from a plume source region. We find that the nighttime and daytime brightness temperatures near Pwyll Crater cannot be matched by including excess heat flow at that location. Rather, we can successfully model both measurements by increasing the local thermal inertia of the surface. 

\end{abstract}

\keywords{planets and satellites: general --- planets and satellites: individual (Europa) --- planets and satellites: surfaces}

\section{Introduction}\label{sec:intro}
Europa may be one of the most habitable worlds in the Solar System. Beneath a relatively thin ice shell, it hosts a liquid water ocean in contact with a rocky core \citep{Anderson1998,Kivelson2000}. Europa's young surface age, surface geology, and salty surface composition point to a history of geologic activity that may have facilitated contact between the ocean and the surface environments \citep[e.g.][]{Bierhaus2009, Kattenhorn2009, McCord1999, Fischer2015}. If such activity continued today, then direct study of the oceanic composition may be possible, but the question of modern geologic activity remains. Recent observations using the Hubble Space Telescope (HST) have hinted at the possibility of active water vapor plumes at Europa \citep{Roth2014, Roth2014b, Sparks2016, Sparks2017}. However, the detections have been sporadic and tenuous, making confirming the existence of plumes difficult. 

It is possible that sites of recent or ongoing geologic activity would cause persistent spatially localized thermal anomalies, similar to the so-called "tiger stripes" of Enceladus \citep{Spencer2006}. Therefore, high-resolution thermal data may present another, perhaps more robust, way of identifying active regions in the case of Europa. 

\citet{Sparks2016} and \citet{Sparks2017} detected potential off-limb absorption near the crater Pwyll in HST images of Europa as it transited Jupiter. This location is coincident with a nighttime thermal excess in brightness temperatures measured by the Galileo Photopolarimeter-Radiometer (PPR) \citep{Spencer1999, Moore2009}. If the thermal excess were caused by increased subsurface heat flow, this association could corroborate the interpretation that the off-limb features are due to subsurface geologic activity and the venting of plume material. However, endogenic heating is not the only potential cause of thermal anomalies; they can also be due to localized variations in surface properties, such as albedo or thermal inertia. In the case of nighttime thermal anomalies, like the one in question, thermal inertia becomes particularly important. Indeed, \citet{Spencer1999} cite thermal inertia as a potential explanation for the nighttime brightness temperatures near Pwyll. 

Anomalies caused by variations in thermal inertia and in endogenic heat flow should have diurnal temperature curves that behave differently over the course of a Europa day, making distinguishing between these explanations possible with temperature measurements at more than one time of day. The published Galileo PPR maps include only a single nighttime observation of the region surrounding Pwyll Crater \citep{Spencer1999}. We present a complementary thermal observation obtained using the Atacama Large Millimeter Array (ALMA), which captures the region of interest during the daytime. Using a thermal model, we fit both the ALMA and Galileo PPR observations and evaluate whether the anomaly is best explained by variation in the thermal inertia or whether it is truly indicative of an endogenic hot spot. 

\section{ALMA Observations and Data Reduction}\label{sec:methods}
The observations described herein were undertaken with the 12-m array of the Atacama Large Millimeter Array (ALMA).  This synthesis array is a collection of radio antennas, each 12 m in diameter, spread out on the Altiplano in the high northern Chilean Andes.  Each of the pairs of antennas acts as a two-element interferometer, and the combination of all of these individual interferometers allows for the reconstruction of the full sky brightness distribution, in both dimensions.

ALMA is tunable in 7 discrete frequency bands, from $\sim$ 90 to $\sim$ 950 GHz.  The observation described in this paper was taken in Band 6, near 230 GHz, in the ``continuum'' (or ``TDM'') mode, with the standard frequency tuning.  For band 6, this yields four spectral windows in the frequency ranges: 224--226 GHz; 226--228 GHz; 240--242 GHz; and 242--244 GHz.  In the final data analysis, we averaged over the entire frequency range in both bands, and we use 233 GHz as the effective frequency in our modeling.

We observed Europa with ALMA on November 27 of 2015 from 10:00 to 10:40 UTC. At the center time of the observation, the sub-Earth longitude was 319.5$\degree$ and the sub-Earth latitude was -1.54$\degree$, capturing Pwyll Crater in the afternoon at $\sim$ 60$\degree$ past local noon. ALMA had 50 available antennas in its C36-7 configuration, with a maximum useable antenna spacing of $\sim$ 5 km.

Fully calibrated visibility data were provided by ALMA. We performed several iterations of self-calibration \citep{Taylor1999} on the visibility data to create a deconvolved Europa image with the 0.05" effective resolution of the interferometer. Figure \ref{fig:alma} shows this image with a pixel sampling of 10 times the full spatial resolution. With this resolution, and given Europa's projected diameter of 0.77" on the sky, we obtained $\sim$15 resolution elements across the disk.

\section{Thermal Modeling}\label{sec:model}
We develop a global thermal diffusion model for Europa, similar to those developed in the past for several solar system bodies \citep[e.g][]{Spencer1989, Spencer1990, Hayne2015}. The model begins with calculations of solar insolation across the surface, where the solar flux absorbed at each point on the disk is given by

\begin{equation}
%\begin{displaymath}
F_{abs}=(1-A)\mu \frac{F_{solar}}{r^2}.
%\end{displaymath}
\end{equation}
Here, $\mu$ is the cosine of the solar incidence angle with respect to the local surface normal, $F_{solar}$ is the solar constant at 1 AU, $r$ is the solar distance in AU, and $A$ is the hemispherical albedo at that point.

In the absence of anomalous endogenic heating and in the limit of very low thermal inertia, the surface temperatures are the result of instantaneous radiative equilibrium with the absorbed flux. However, real bodies will have a finite thermal inertia,

\begin{equation}
%\begin{displaymath}
I=\sqrt{\rho c_pK},
%\end{displaymath}
\end{equation}
where $\rho$ is the density, $c_p$ is the specific heat capacity, and $K$ is the thermal conductivity, resulting in a diurnal thermal wave with depth. Temperature as a function of time, t, and depth, z, is then given by the one-dimensional heat equation

\begin{equation}
%\begin{displaymath}
\rho c_p \frac{\partial{T}}{\partial{t}} = \frac{\partial}{\partial{z}}\left(K\frac{\partial{T}}{\partial{z}}\right).
%\end{displaymath}
\end{equation} 
The model achieves a numerical solution to this equation by computing finite differences across depth elements. We assume a global heat flux of 20 $mW/m^2$ \citep{MitriShowman2005, BarrShowman2009} as a lower boundary condition and an outgoing surface flux of $\epsilon \sigma T^4$ as an upper boundary condition, where $\epsilon$ and $\sigma$ are the bolometric emissivity and Stefan-Boltzmann constant, respectively. We simulate a total of 5 diurnal skin depths, where the skin depth is given by

\begin{equation}
%\begin{displaymath}
d=\sqrt{\frac{2KP}{\rho c_p}}
%\end{displaymath}
\end{equation} 
and P is the rotational period of Europa. We define the thickness of the top layer to be $d/30$, with the thickness of each subsequent layer increasing by a factor of 1.2.

We run the model using a time step no greater than 1/500 of a Europa day and allow it to equilibrate until the maximum difference in surface temperature across 5 Europa days at any point on the disk is less than 1$\%$ (usually 10 - 15 Europa days). Further equilibration results in differences $\ll$1 K everywhere on the disk. The product is a temperature map of the surface of Europa, which can be output at any point in time throughout the Europa day.

We use this simple thermal model combined with an approximated high-resolution albedo map of the surface to simulate the ALMA and Galileo observations and establish a baseline against which to assess thermal anomalies. Since no published high-resolution albedo map of the surface exists, we construct a high-resolution map by using discrete Voyager normal albedos as tie-points to the USGS Voyager/Galileo greyscale basemap of Europa (at 2 pixel/degree resolution) \citep{USGSmap}. We take the normal albedo points of \citet{McEwen1986} in the Voyager green, blue, violet, and ultraviolet filters, weight them by the width of each filter and the magnitude of the solar flux at the relevant wavelengths, and then add them to get approximate wavelength-integrated normal albedos. We then find the greyscale value of each corresponding point in the USGS basemap and use the resulting linear correlation to get an approximate wavelength-integrated normal albedo for each point in the USGS map. The greyscale values and approximate normal albedos correlate linearly with an $R^2$ of .92 and a standard deviation of .03, which we take as the statistical error in our albedos. As the phase integral of Europa is 1.01 \citep{Grundy2007}, we take these normal albedos as approximate hemispherical albedos and use them in model calculations of solar flux absorbed across the surface of Europa. 

In modeling the observations, we take the thermal inertia and emissivity as free parameters and treat the entire disk as homogenous in these properties. We assume a snow-like constant regolith density of 500 $kg/m^3$ \citep{Spencer1999} and a $c_p$ of 900 $J/(K \cdot kg)$, which is appropriate for water ice near 100 K \citep{FeistelWagner2006}. After equilibration, we halt the simulation at the time specified by the sub-solar longitude of the observation. We then convert surface temperatures into flux units via Planck's Law and project the model output based on the viewing geometry of the observation, such that the central point on the disk corresponds to the sub-observer coordinates. 

When modeling the ALMA observation, we apply a Gaussian filter with full widths at half maximum (FWHMs) corresponding to those of the elliptical ALMA beam to smooth the output to match ALMA resolution. When modeling the Galileo PPR observation, we apply a Gaussian filter consistent with a 140-km linear resolution, within the 80--200-km resolution of the PPR observations \citep{Spencer1999}. Finally, we convert the smoothed images into brightness temperature, again using Planck's Law, and compare them to the actual observations. The nighttime PPR observation of Pwyll was taken in the open filter position (sensitive from 0.35 to $\sim$ 100 $\micron$), but these brightness temperatures generally agreed to $<$ 1 K with those taken in the 27.5 $\micron$ filter \citep{Spencer1999}. Thus, in modeling the PPR observation, we output brightness temperatures for a wavelength of 27.5 $\micron$ \citep{Spencer1999}. In modeling the ALMA observation, we calculate brightness temperatures at a wavelength of 1.3 mm (233 GHz). We treat the emissivities of the surface at the ALMA and PPR wavelengths as equal. This assumption is reasonable in the case of water ice under laboratory conditions, as the optical constants are similar at both wavelengths \citep{Warren2008}. However, the relevant emissivities for Europa-like conditions and compositions are not known.

It should be noted that, for emissivities less than 1, the resulting brightness temperatures at these two wavelengths will be significantly different. While the ALMA wavelength is nearly in the Rayleigh-Jeans limit, the Galileo PPR was sensitive to Europa's $\sim$ 30 $\micron$ blackbody peak. In both cases, Planck's law can be used to find the brightness temperature in terms of the physical temperature

\begin{equation}
%\begin{displaymath}
T_b = \frac{h\nu}{k}\left(ln\left[1+\frac{e^{h\nu/kT}-1}{\epsilon}\right]\right)^{-1},
%\end{displaymath}
\end{equation} 
where $\nu$ is the frequency, h is Planck's constant, k is Boltzmann's constant, and $\epsilon$ is the emissivity. This equation gives different results for the two wavelength regimes. For instance, an emissivity of 0.8 and a physical temperature of 125 K, produce a brightness temperature of 119 K at 27.5 $\micron$ and a brightness temperature of 101 K at 1.3 mm. It should also be noted that the two observations were taken at very different solar distances. During the PPR observation, Jupiter was near perihelion (at 4.96 AU). However, it was near aphelion (at 5.4 AU) in 2015 when the ALMA image was taken. We account for this effect in our model.

Some caveats do apply to our very simple model, however. First, we do not include the effects of surface roughness. Rough topography has the tendency to enhance surface temperatures, with the largest effects appearing at the limbs. However, Europa is thought to be relatively smooth compared to other solar system bodies \citep[e.g.][]{Spencer1987,Domingue1997}. We tested a roughness model with rms slopes up to 20$\degree$, using a similar implementation of surface roughness to \citet{Hayne2015}, and found that the effects did not significantly affect our results. 

Second, our model assumes that the thermal emission imaged in the ALMA and Galileo observations originates from the topmost model layer. For the Galileo PPR observations, which were sensitive to the $\sim$ 30 $\micron$ blackbody peak of Europa, this is a valid assumption. However, ALMA senses slightly deeper into the surface at a wavelength of 1.3 mm. Thus, model ALMA brightness temperatures for a given $\epsilon$ and $I$ are slightly warmer than they would be if this effect were included. In testing a variation of our simple model, which included sensing beneath the surface with an e-folding of 1 cm, we found that much of this brightness temperature variation was captured by a slight change in the model emissivity, $\epsilon$.

Finally, our model assumes that all of the absorbed solar flux is captured in the topmost layer. This is the standard assumption in many thermal models \citep[e.g.][]{Spencer1989,Spencer1990,Hayne2015}, and is valid for solar system bodies with low bolometric albedos. However, it is possible that sunlight is able to penetrate to significant depths beneath a high albedo surface, such as that of much of Europa \citep[e.g][]{Brown1987, Urquhart1996}. As this effect can create a heat reservoir at depth, it can be difficult to distinguish from a change in thermal inertia \citep{Urquhart1996}. We found this to be true in testing a version of our model that also included sunlight propagation with an e-folding of 2 cm, and this effect did not improve our fits to the data. 

For this analysis, we are primarily interested in relative local variation in the thermal parameters near Pwyll Crater, rather than in accurately determining the true global values. Thus, we choose to present the simplest model, with the knowledge that some of the caveats discussed here may manifest as changes in our model parameters. 

\section{Fits to ALMA and Galileo PPR Observations}\label{sec:fits} 
\begin{figure*}
\figurenum{1}
\plotone{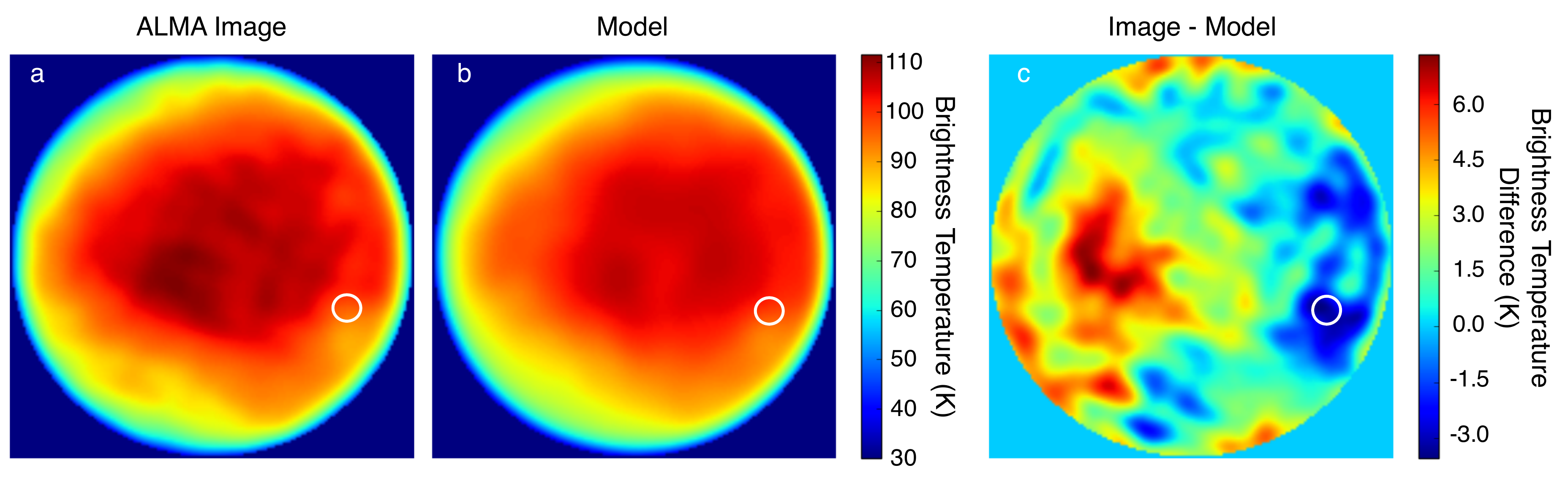}
\caption{Model fit to the ALMA data. (a) shows the ALMA image in brightness temperature at $\lambda$ = 1.3 mm. (b) shows our model image using the best-fit parameters to both the ALMA and PPR data. (c) shows the residuals between the model and the data, where positive values indicate locations where the data are warmer than the model. The location of the potential plume source region and Galileo thermal anomaly is circled in white, where the size of the circle corresponds to the size of our ALMA resolution element. \label{fig:alma}}
\end{figure*}

\begin{figure*}
\figurenum{2}
\plotone{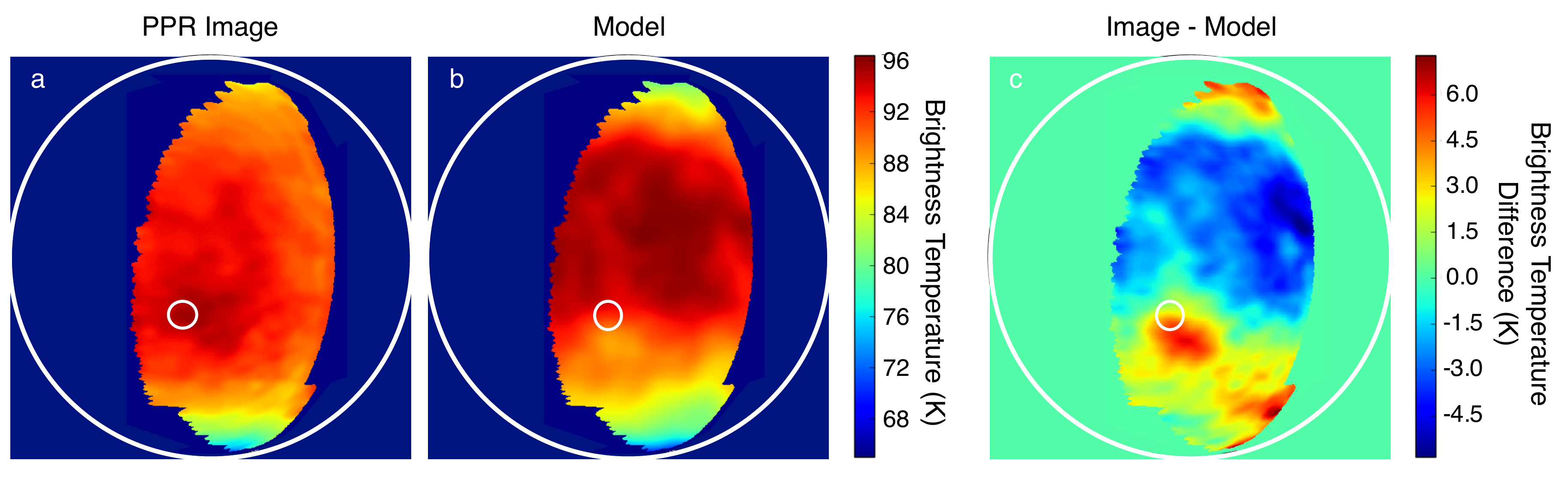}
\caption{Model fit to the PPR data. (a) shows the PPR image in brightness temperature at $\lambda$ = 27.5 $\micron$. (b) shows our model image using the best-fit parameters to both the ALMA and PPR data. (c) illustrates the residuals between the model and the data, where positive values indicate locations where the data are warmer than the model. The location of the potential plume source region and Galileo thermal anomaly is circled in white, where the size of the circle corresponds to the size of our ALMA resolution element. \label{fig:ppr}}
\end{figure*}

In order to obtain simultaneous best-fit parameters to both the ALMA and Galileo PPR data, we run our model for each observation over a wide grid of thermal inertias and emissivities and minimize the sum of the squares of the residuals for the region covered by both datasets. By fitting both observations at once, we find that an emissivity of 0.8 and a thermal inertia of 95 $J/(m^2 \cdot K \cdot s^{1/2})$ provide the best result for the overlapping region. This thermal inertia is slightly higher than the value of 70 $J/(m^s \cdot K \cdot s^{1/2})$ reported by \citet{Spencer1999} for the equatorial latitudes, but is within the range of 30--140 $J/(m^s \cdot K \cdot s^{1/2})$ calculated by \citet{Rathbun2010}. Our ALMA observation, best-fit model ALMA image, and the corresponding residuals are shown in Figure \ref{fig:alma}, where the approximate location of the Galileo thermal anomaly and potential plume source region is circled. While the homogenous thermal model is able to reproduce the large-scale structure of the ALMA image well, there are significant localized discrepancies, which are not necessarily surprising given the inhomogeneous nature of Europa's surface. This observation includes much of the dark trailing hemisphere of Europa, which is compositionally diverse \citep{Carlson2009}. Therefore, we do not expect the surface to be well-represented by a single thermal inertia or emissivity. However, it is interesting that the area associated with the Galileo nighttime thermal excess is actually colder in the ALMA data than the model predicts. For the Galileo PPR observation, shown in Figure \ref{fig:ppr} alongside the best-fit model image and the resulting residuals, the same location is indeed anomalously warm, as noted by \citet{Spencer1999} and \citet{Sparks2017}. In fact, the entire Pwyll Crater region, not just the potential plume source location slightly northwest of the crater, shows up as anomalously hot at night and cold during the day. This pattern is suggestive of a variation in the local thermal inertia. If the thermal anomaly were instead due to an endogenic hot spot with an excess subsurface heat flux, one would expect the area to have elevated brightness temperatures throughout the diurnal cycle. 

\begin{figure*}
\figurenum{3}
\plotone{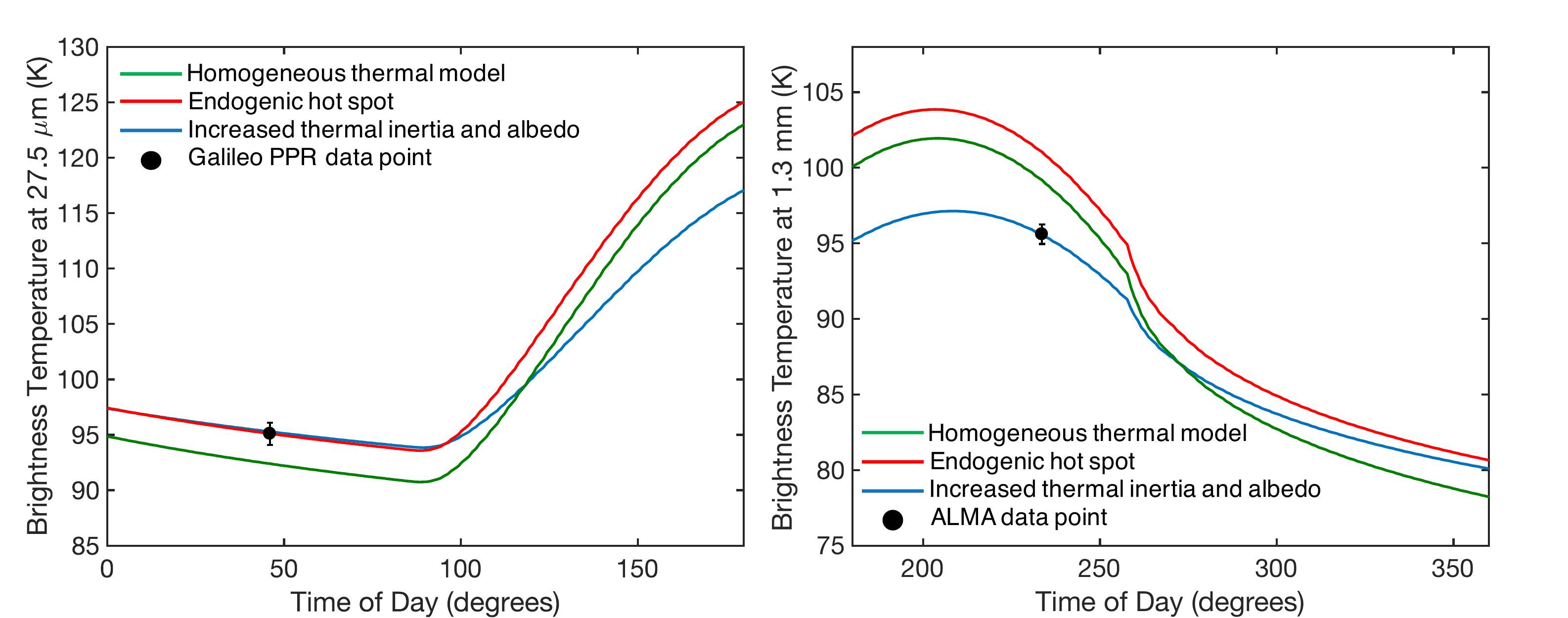}
\caption{Model fits to the ALMA and PPR data points. Here, 180$\degree$ indicates local noon and 0$\degree$ is local midnight. Both points are fit well by raising the thermal inertia and adjusting the albedo by $\sim$ 5$\%$. Invoking an endogenic hotspot to explain the Galileo nighttime thermal anomaly, however, results in a daytime brightness temperature much higher than that measured by ALMA. The homogenous thermal model obtained by simultaneously fitting the Galileo and ALMA observations fails to fit the anomalous region during either the night or the day. The steep drop in brightness temperature at $\sim$ 260$\degree$ coincides with Europa being eclipsed by Jupiter. Also, note the differences in the y-axes. The Galileo point is fit with brightness temperatures at 27.5 $\micron$, while the ALMA point is fit with brightness temperatures at 1.3 mm. The noise levels of the PPR observations are stated to be $<$ 1 K \citep{Spencer1999}, and a 1 K error bar is shown here. The statistical error on the ALMA measurement is +/- 0.6 K. \label{fig:curves}}
\end{figure*}

To investigate whether the ALMA and PPR brightness temperatures are best explained by an endogenic hot spot or a thermal inertia anomaly, we model the location of the anomaly under both scenarios over the course of a diurnal cycle and attempt to fit both data points. We simulate an area 156 km in radius (corresponding to our ALMA resolution in this region) centered on 276$\degree$ W and 16.8$\degree$ S, which is coincident with the Galileo thermal anomaly in the potential plume source region \citep{Sparks2016, Sparks2017}. We model the case of an endogenic thermal anomaly by raising the geothermal heat flux beneath the lowest layer of our simulation (5 diurnal skin depths $\approx$ 0.75 m at our best-fit parameters). We define the best fits by minimizing the sum of the squares of the differences between the models and the two data points. The results of these fits are shown in Figure \ref{fig:curves}, where the ALMA data point is taken to be 95.6 K, the ALMA brightness temperature at 276$\degree$ W and 16.8$\degree$ S, and the PPR data point is 95.1 K, the brightness temperature given by averaging the measured flux over an area 156 km in radius centered on the same location.

Overall, we find that the ALMA and Galileo measurements are best explained by invoking a thermal inertia anomaly, and that the anomalous region cannot be solely attributed to endogenic heating. We successfully match both measurements by increasing the thermal inertia by 47$\%$ from 95 to 140 $J/(m^2 \cdot K \cdot s^{1/2})$ and increasing the albedo of the region by 5$\%$ from 0.56 to 0.59, which is within our albedo uncertainties. However, we are unable to successfully fit both brightness temperatures with a subsurface hot spot. Reproducing the Galileo nighttime brightness temperature requires raising the subsurface heat flux from 0.02 $W/m^2$ to 0.66 $W/m^2$, which produces a daytime brightness temperature much higher than we observed with ALMA (Figure \ref{fig:curves}).

Similarly, a combination of subsurface heating and a thermal inertia anomaly cannot explain the two measurements. As endogenic heating increases both day and night temperatures, fitting one of the two data points in this manner always overestimates the brightness temperature of the other. We can only invoke endogenic heating in matching the data if we also significantly raise the local albedo. A local heat flux of 1.6 $W/m^2$ can account for both the ALMA and Galileo PPR brightness temperatures, but only when combined with a local albedo increase of 23$\%$ from 0.56 to 0.69, a 4$\sigma$ deviation from our albedo model. Discrepancies of this magnitude would only result from systematic biases, rather than occur in isolation at one location. Systematic albedo biases would affect the entire albedo map and be largely absorbed by changes in the best-fit parameters. Thus, we argue that the simplest and most likely explanation for the Galileo nighttime thermal anomaly near Pwyll Crater is a moderate increase in the local thermal inertia.

Spatially localized thermal inertia variations can result from a number of causes, including compositional differences and changes in the average grain size of the surface material. An elevated thermal inertia near Pwyll Crater and the anomaly in question, as originally noted by \citet{Spencer1999}, may result from higher average regolith particle sizes in the ejecta blanket. This possibility seems particularly plausible as the anomalous temperatures are not just constrained to the relatively small potential plume source area \citep{Sparks2017}, but are observed across the entirety of the Pwyll region (Figures \ref{fig:alma} and \ref{fig:ppr}). \citet{Spencer1999} also suggest the possibility that impact-exposed water ice may allow for deeper sunlight penetration, which, as discussed in Section \ref{sec:model}, can mimic the effects of increased thermal inertia.

One final potential explanation warrants mentioning. We cannot rule out the possibility that the region associated with the thermal anomaly was anomalously warm due to endogenic heat at the time of the Galileo observation in 1998, but has since cooled. For instance, if a hot spot were not actively heated, but rather were caused by a singular upwelling of liquid water or warm ice at or near the surface, then detectable heat signatures need not necessarily last the 17 years between the Galileo and ALMA observations \citep{Abramov2008}. However, our ALMA observation was taken in 2015, prior to the 2016 potential plume detection of \citet{Sparks2017}. Thus, if the hot spot had dissipated by the time of our observation, then the same anomaly cannot be linked to that of \citet{Sparks2017}.

\section{Conclusions}\label{sec:conclusions}
Using ALMA, we obtained a daytime thermal measurement of the Galileo PPR nighttime thermal excess \citep{Spencer1999} near Pwyll Crater, which is associated with two potential plume detections \citep{Sparks2016, Sparks2017}. If the thermal excess were due to an endogenic hot spot, then it could support the idea that the region northwest of Pwyll exhibits modern geologic activity. Using a global one-dimensional thermal diffusion model, we fit both the ALMA and PPR observations. However, while the location in question does appear hot relative to our model at night, it appears colder in our ALMA daytime image than the model predicts. We suggest that this pattern is indicative of a locally elevated thermal inertia. To investigate whether we can simultaneously explain both temperature measurements with endogenic heating or need to invoke a thermal inertia anomaly, we model the potential plume source location over the entire course of a Europa day under both scenarios and attempt to fit the two measured brightness temperatures. While we can explain the Galileo nighttime brightness temperature with an endogenic heat source, this situation results in a daytime brightness temperature that is too hot. However, we successfully fit both observations by raising the local thermal inertia by 47$\%$ and adjusting the albedo by an amount within our uncertainties. We therefore conclude that the nighttime Galileo thermal anomaly is most likely explained by a variation in the local surface thermal inertia, which may result from its proximity to the crater Pwyll.

\acknowledgements
This paper makes use of the following ALMA data: ADS/JAO.ALMA\#2015.1.01302.S. ALMA is a partnership of ESO (representing its member states), NSF (USA) and NINS (Japan), together with NRC (Canada), MOST and ASIAA (Taiwan), and KASI (Republic of Korea), in cooperation with the Republic of Chile. The Joint ALMA Observatory is operated by ESO, AUI/NRAO and NAOJ. The National Radio Astronomy Observatory is a facility of the National Science Foundation operated under cooperative agreement by Associated Universities, Inc. This research was supported by Grant 1313461 from the National Science Foundation. The authors thank John R. Spencer for kindly providing the Galileo PPR data used within this paper.

\end{document}